\documentclass[aps,pra,pdf,superscriptaddress,amsmath,amssymb,amsfonts,twocolumn,showpacs,nofootinbib,longbibliography]{revtex4-1}

\usepackage{amssymb}
\usepackage{eqnarray,amsmath}
\newcommand{\abs}[1]{\left| #1 \right|} 

\usepackage{graphicx}
\usepackage{epsfig}
\usepackage{dcolumn}
\usepackage{bm}
\usepackage{braket}
\usepackage{amsmath}
\usepackage{hyperref}
\usepackage{physics}
\usepackage{mathtools}
\usepackage{graphicx,color,xcolor,colortbl}

\usepackage[normalem]{ulem}

\definecolor{Gray}{gray}{0.85}
\definecolor{LightCyan}{rgb}{0.88,1,1}

\newcolumntype{a}{>{\columncolor{Gray}}c}
\newcolumntype{b}{>{\columncolor{white}}c}

\usepackage{hyperref}

\hypersetup{colorlinks=true, citecolor=blue, urlcolor=blue, linkcolor=blue}

\begin{document}

\title{Tunneling dynamics of $^{164}$Dy supersolids and droplets} 

\author{S. I. Mistakidis}
\affiliation{Department of Physics, Missouri University of Science and Technology, Rolla, MO 65409, USA}
\affiliation{ITAMP, Center for Astrophysics $|$ Harvard $\&$ Smithsonian, Cambridge, MA 02138 USA}
\affiliation{Department of Physics, Harvard University, Cambridge, Massachusetts 02138, USA} 
\author{K. Mukherjee}
\affiliation{Mathematical Physics and NanoLund, Lund University, Box 118, 22100 Lund, Sweden}
\author{S. M. Reimann}
\affiliation{Mathematical Physics and NanoLund, Lund University, Box 118, 22100 Lund, Sweden} 
\author{H. R. Sadeghpour}
\affiliation{ITAMP, Center for Astrophysics $|$ Harvard $\&$ Smithsonian, Cambridge, MA 02138 USA}

\date{\today}

\begin{abstract}
The tunneling dynamics of a magnetic $^{164}$Dy quantum gas in an elongated or pancake skewed double-well trap is investigated with a time-dependent extended Gross-Pitaevskii approach. 
Upon lifting the energy offset, different tunneling regimes can be identified.
In the elongated trap and for sufficiently large offset, the different configurations exhibit collective macroscopic tunneling. 
For smaller offset, partial reflection from and transmission through the barrier 
lead to density accumulation in both wells, and eventually to tunneling-locking. One can also reach the macroscopic self-trapping regime for increasing relative dipolar interaction strength, while tunneling vanishes for large barrier heights. 
A richer dynamical behavior is observed for the  pancake-like trap. 
For instance, the supersolid maintains its shape, while the superfluid density gets distorted signifying the emergence of peculiar excitation patterns in the macroscopic tunneling 
regime. The findings reported here may offer new ways to probe distinctive dynamical features in the supersolid and droplet regimes.
\end{abstract}

\maketitle
	
\section{Introduction}\label{Intro}

Dipolar Bose-Einstein condensates (dBECs) consisting of magnetic chromium~\cite{Griesmaier2005} or lanthanides such as dysprosium~\cite{Lu2011}, erbium~\cite{Aikawa2012} and  europium~\cite{Miyazawa2022}  offer highly flexible platforms to unravel exotic many-body phases of matter~(for reviews, see Refs.~\cite{Lahaye2009,Bottcher2020,Chomaz2023}). 
The interplay of long-range anisotropic dipole-dipole interactions (DDI), short-range isotropic interactions and quantum fluctuation contributions  may prevent a collapse of the dBEC~\cite{Kadau2016,Ferrier2016,Chomaz2016,Schmitt2016,Wachtler2016a,Wachtler2016b}. It gives rise to the -- by now well-studied -- supersolid (SS) and droplet phases~(see the recent review, Ref.~\cite{Chomaz2023}). 

Advances in ultra-cold atoms have opened up new possibilities to observe this intricate SS state of matter.  
Predictions were made for Rydberg-excited BECs~\cite{Henkel2010} and in the dipole-blockade regime~\cite{Cinti2010}. Early experimental realizations of supersolidity were reported for example in spin-orbit~\cite{Li2017} and BECs coupled to optical cavities~\cite{Leonard2017a,Leonard2017b}.  
The above mentioned dBECs, however, provide a particularly favorable setting since the localization is not externally initiated, 
but rather is an intrinsic property resulting from the spontaneous symmetry breaking due to the interactions. 
Only shortly after the discovery of dipolar droplets, the SS state was uncovered in three milestone experiments 
with highly magnetic dysprosium and erbium atoms~\cite{Bottcher2019,Tanzi2019a,Chomaz2019}.

In dBECs, for the SS and self-bound states, quantum fluctuation plays a major role. 
Similar to the description of stable self-bound droplets in binary 
BECs~\cite{Petrov2015}, they are commonly modelled with the extended Gross-Pitaevskii method (eGPE)~\cite{Wachtler2016a,Wachtler2016b,Bisset2016,Chomaz2016,Baillie2018}, including the Lee-Huang-Yang (LHY) correction~\cite{Lee1957,Lima2011} to approximate quantum fluctuations %higher-order corrections 
to first order. 
In a dipolar SS, the spontaneous breaking of translational symmetry~\cite{Bottcher2019,Tanzi2019a,Chomaz2019,Norcia2021a} gives rise to density modulations while partially retaining superfluid (SF) properties and exhibits both diagonal and off-diagonal long-range order~\cite{Gross1957,Gross1958,Yang1962,Andreev1969,Chester1970,Leggett1970}. 
This phenomenon is underpinned by the presence of a roton minimum in the energy-momentum dispersion relation which has been experimentally probed~\cite{Chomaz2018,Guo2019,Natale2019}.  Notably, in cases where long-range interactions are dominant, the dilute SF background density, characteristic of the SS phase, may vanish leading to the formation of isolated droplet lattice (DL) configurations~\cite{Ferrier2016,Bottcher2019,Chomaz2016}. 

SS and DL phases have been realized in elongated traps, where they can order periodically along one spatial dimension~\cite{Bottcher2019,Chomaz2019,Tanzi2019a,Tanzi2019b} and more recently in planar droplet arrays in two dimensions~\cite{Norcia2021a,Poli2021,Bland2022a}.  In oblate trap geometries, states of dysprosium with different ground state morphologies were predicted, from honeycomb to triangular, striped and ring-shaped lattices~\cite{Zhang2021,Hertkorn2021c,Schmidt2022}.
In the presence of an optical lattice, a geometrical frustration of the dipolar droplet lattice may be induced, and more complicated phase patterns emerge~\cite{Halperin2023}. 
Dipolar mixtures have more recently been investigated~\cite{Smith2021,Li2022,Politi2022,Scheiermann2023}, 
for which alternating-domain supersolids have also been found~\cite{Bland2022b}. 
Dynamical properties offer intriguing future prospects~\cite{Norcia2021b}, 
and notable examples include (but are not limited to) interaction quenches~\cite{Tanzi2019a,Bottcher2019,Halder2022,Mukherjee2023a},  the formation of vortex configurations~\cite{Ancilotto2021,Klaus2022,Bland2023}, the manifestation of persistent currents~\cite{Tengstrand2021,Tengstrand2023}  or the impact of 
relevant thermal effects~\cite{Sohmen2021,Bland2022a,Sanchez2023}. 

For BECs with short-range interactions, the tunneling dynamics in double-well potentials has been widely investigated, see for instance the review~\cite{Gati2007} or the collected articles  in~\cite{Keshavamurthy2011}. 
Such systems constitute atomic analogues of the well-known superconducting Josephson junction~\cite{Josephson1962} and  exhibit tunneling phenomena~\cite{Albiez2005,Gati2007} including, for instance, Josephson oscillations~\cite{Smerzi1997} and macroscopic quantum self-trapping~\cite{Raghavan1999}. 
However, the tunneling dynamics of dBECs in the SS and DL phases and in particular, the interplay of short- and long-range interactions on the emergent tunneling regimes remain largely unexplored. 
Additionally, since the droplet crystal configuration depends on the spatial dimension of the system, it would be intriguing to explore the impact of dimensionality on the emergent tunneling behavior. Relevant open questions, for instance, whether the crystal arrangement retains its symmetry in the course of tunneling and how suppression of the latter occurs due to interactions in various dimensions, are addressed. 

Here, we report an attempt to theoretically investigate the tunneling dynamics of a dipolar SS or DL in a double-well potential
in the eGPE framework. The approach is anticipated to capture the basic dBEC tunneling dynamics, whilst higher-order effects, such as additional tunneling channels or inter-band contributions,  require more sophisticated approaches~\cite{Mistakidis2022,wilsmann2018control} and are left for future work. 

The dBEC is initially confined in a tilted double-well potential, with the individual wells chosen to be elongated or pancake-like. Then the system is suddenly quenched from an initial energy offset between the wells to a symmetric double-well potential. 
The initial energy offset, the relative dipolar strength ({\textit {i.e.}}, contact versus dipolar coupling) and the barrier height determine the tunneling characteristics of the system. Prior to the quench, the 
single-well configurations exhibit the well-known formation of SF, SS, and DL  states~(see {\textit e.g.} the review of  Ref.~\cite{Chomaz2023}).
For an elongated double-well with a sufficiently large initial energy offset, the different dBEC configurations exhibit periodic oscillations between the wells with constant amplitude and frequency, thus featuring collective tunneling. 
When the offset is reduced, partial reflection and transmission occur through the barrier, resulting in the accumulation of density in both wells and eventually leading to tunneling-locking.  
At long evolution times, in the self-bound DL regime, a significant population difference is established, distinguishing it from the SF phase where the imbalance vanishes. Additionally, the SS and DL states exhibit a reduced center-of-mass velocity compared to the SF phase, indicating their rigidity. Increasing the barrier height while keeping all other parameters fixed, it is possible to enter the macroscopic self-trapping regime.
The different patterns also manifest in pancake-like double-well systems, however with a richer phenomenology. 

This paper proceeds as follows: Section~\ref{theoy_sec} introduces the dBEC setting  and the relevant eGPE framework. The ground-state phases of the system confined in a double-well with an energy offset for varying interactions are discussed in Sec.~\ref{initialization}. 
Consecutively, the resultant tunneling dynamics of the different phases after suddenly switching off the underlying energy offset is analyzed for an elongated double-well in Sec.~\ref{tunnel1D} and pancake-like  one in Sec.~\ref{tunnel2D}. 
We summarize and comment on future perspectives in Sec.~\ref{conclusion}. 
In Appendix~\ref{3body}, we discuss the effect of three-body recombination on the tunneling behavior, while Appendix~\ref{td_quench} explicates the persistence of tunneling after a linear ramp of the energy offset. 
Appendix~\ref{numerics} further elaborates on some technical details.

\section{Extended Gross-Pitaevskii for the dipolar gas}\label{theoy_sec}

We consider a dBEC of $^{164}$Dy atoms having a magnetic dipole moment $\mu_{m}=9.93\mu_{B}$ (where $\mu_{B}$ is the Bohr magneton) being polarized along the $z$-direction. 
At zero temperature, the dynamics of the dBEC is modeled by the eGPE~\cite{Wachtler2016a,Wachtler2016b,Bisset2016,Chomaz2016,Baillie2018}  
containing the first order LHY beyond mean-field correction, 
\begin{eqnarray}\label{eGPE}   
& i\hbar \frac{\partial \psi(\textbf{r},t)}{\partial t}  =  \bigg[-\frac{\hbar^2}{2m}\nabla^2 + V(\textbf{r}) + \frac{4\pi\hbar^2 a_s}{m}  \abs{\psi(\textbf{r},t)}^2+\nonumber\\& \gamma(\epsilon_{{\rm dd}})\abs{\psi(\textbf{r},t)}^3 +\int dr^{\prime} U_{dd}(\textbf{r-r}^{\prime})\abs{\psi(\textbf{r}^{\prime},t)}^2 \bigg] \psi(\vb{r},t). 
\end{eqnarray}
The three-dimensional (3D) wave function is represented by $\psi (\textbf{r} , t)$ and $m$ is the mass of Dy atom. 
The long-range anisotropic DDI is $U_{dd} (\textbf{r},t) = \frac{\mu_0 \mu^2_{m}}{4\pi} \left[\frac{1-3\cos^2\theta}{\vb{r}^3}\right]$. The angle between the relative distance $\vb{r}$ of two dipoles and the $z$ axis of the quantization axis (defined by the magnetic field) is denoted by $\theta$, and $\mu_0$ represents the permeability of the vacuum. 
Moreover, the dipolar atoms experience short-range contact interactions whose strength is quantified via the 3D $s$-wave scattering length, $a_s$, that can be experimentally tuned through Fano-Feshbach resonances~\cite{Koch2008,Ilzhofer2021}. 
The LHY term~\cite{Lee1957} approximates the lowest-order quantum fluctuation contributions to the energy functional. These fluctuations in 3D appear to be  repulsive and scale with the gas density as $\sim n^{3/2}$, taking the form $\gamma(\epsilon_{{\rm dd}}) = \frac{32}{3}g \sqrt{\frac{a_s^3}{\pi}} \left(1+\frac{3}{2}\epsilon_{{\rm dd}}^2\right)$~\cite{Lima2011}. 
Its contribution is crucial for the system to sustain many-body self-bound states such as droplets and SS and it has been shown to adequately describe experimental observations~(as reviewed in~\cite{Chomaz2023}). 

In a 3D harmonic trap the relative strength $\epsilon_{{\rm dd}} = a_{{\rm dd}}/a_s$ between the DDI and the short-range interactions determines the many-body phase of the system~\cite{Ferrier2016}. In particular,  the dipolar length of $^{164}$Dy atoms  is $a_{{\rm dd}}= \mu_0 \mu^2_{m} m /12\pi\hbar^2=131a_{B}$, with $a_{B}$ being the Bohr radius. 
For sufficiently small values of $\epsilon_{{\rm dd}}$, the system exhibits a SF phase. The equilibrium solution is determined by the delicate balance between the attractive or repulsive long-range dipolar interaction and the repulsive contact interaction. As the short-range scattering length decreases, the relative strength of the long-range dipolar interaction becomes  dominant. Within a small range of $\epsilon_{\rm dd}$, the system favors the SS phase, where a  periodic structure of localized densities coherently connected by a dilute superfluid background emerges. However, a further increase in $\epsilon_{{\rm dd}}$ leads to a DL phase.

The initial state is prepared in an external 3D tilted double-well potential,
\begin{equation}
V(\textbf{r}) = \frac{1}{2}m(\omega_x^2 x^2 + \omega_y^2 y^2 + \omega_z^2 z^2)+V_D e^{-\frac{x^2}{2 w^2}}+Dx,\label{double_well}    
\end{equation}
where $V_D$ is the barrier height  and $w$ is the barrier width. 
For our purposes, we use $V_D=10 \hbar \omega_x$ and $w=0.5 l_{{\rm osc}}$, with $l_{{\rm osc}}=\sqrt{\hbar/(m \omega_x)}$ being the harmonic oscillator length and $\omega_x$ denoting the frequency of the external confinement. 
The last term is the linear external field gradient of strength $D$ that energetically favors the population on the left well. 
The tilt strength ranges from $D=1 \hbar \omega_x/l_{{\rm osc}}$ to $D=10\hbar \omega_x/l_{{\rm osc}}$. 

\begin{figure} 
\centering
\includegraphics[width = 0.48\textwidth]{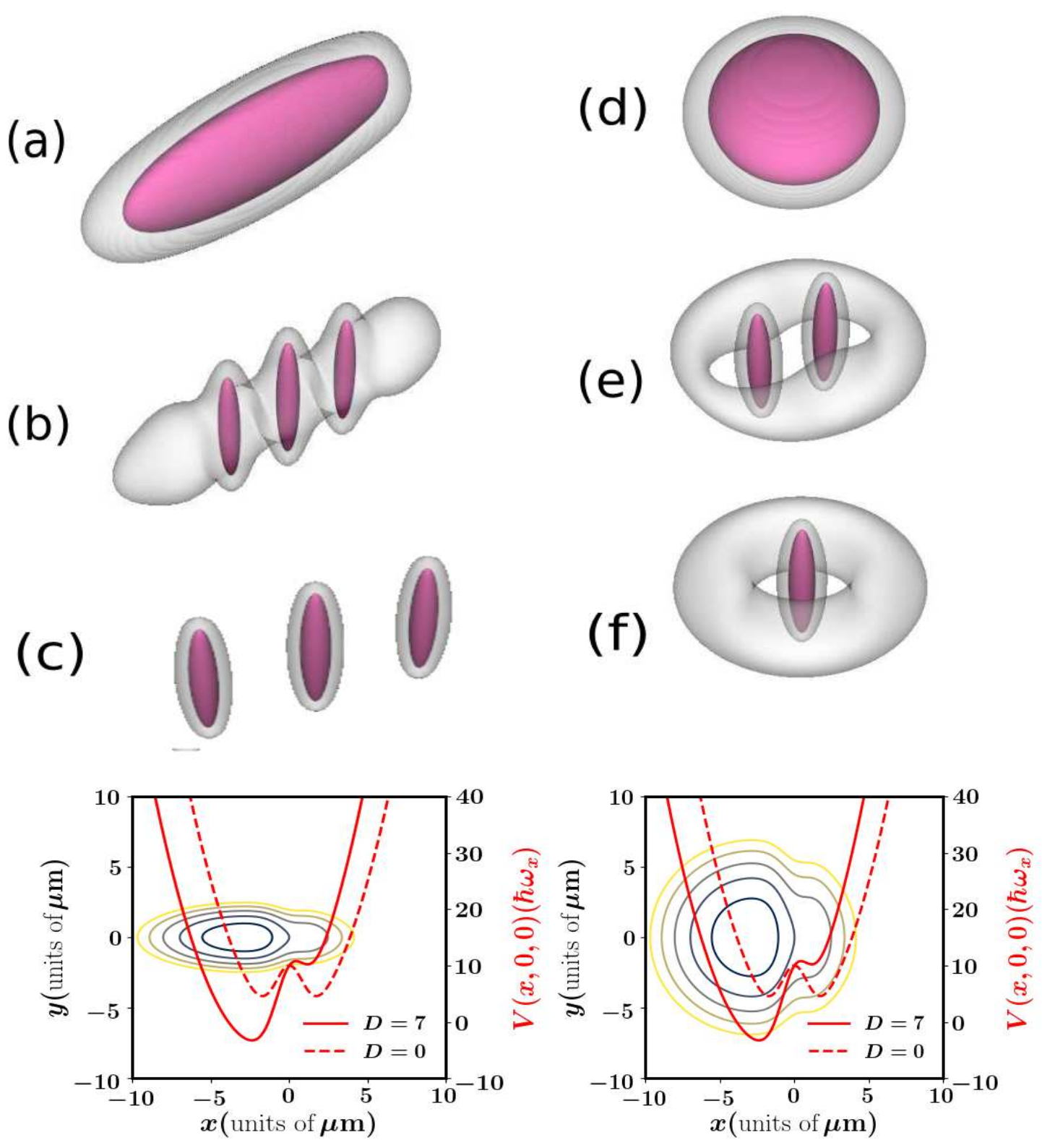}
\caption{Ground state density iso-surfaces of a dBEC initially localized on the left side of (a)-(c) an elongated and (d)-(f) pancake-like tilted 
double-well trap, as depicted by the contour lines in the lowest  panels. The density isosurfaces represent 20\% and 2.5\% of maximum density. 
All atoms reside in the left part of the double-well, and the entire density profile is located in the minimum of the left well. 
Depending on the relative interaction strength $\epsilon_{ {\rm dd}}$ the dBEC in the left well features (a), (d) a regular SF for $\epsilon_{dd}=1.31$, (b) an elongated ($\epsilon_{ {\rm dd}}=1.4$)  or (e) planar SS  ($\epsilon_{ {\rm dd}}=1.49$), (c) an elongated DL pattern for $\epsilon_{dd}=1.49$ as well as (f) a single droplet surrounded by a homogeneous density distribution for $\epsilon_{ {\rm dd}}=1.559$. 
Notice that the individual transitions are shifted to larger $\epsilon_{{\rm dd}}$ from the quasi-1D to the quasi-2D setting. 
The tilt strength is $D=7\hbar \omega_x/l_{{\rm osc}}$ enforcing all $N=40000$ magnetic atoms to reside on the left part of the double-well. 
The latter is characterized by frequencies (a)-(c) $(\omega_x, \omega_y, \omega_z)=2\pi \times (19, 53, 81)\rm Hz$ and (d)-(f) $(\omega_x, \omega_y, \omega_z)=2\pi \times (43, 43, 131)\rm Hz$, while having a width $w=0.5l_{{\rm osc}}$ and height $V_{D}=10\hbar \omega_x$. {The potential contours at $z=0$ plane for $D=7\hbar \omega_x/l_{{\rm osc}}$, along with the variation of the potentials $V(x, 0, 0)$ across $x$ for  $D=7\hbar \omega_x/l_{{\rm osc}}$ (solid line) and $D=0$ (dashed line),  are shown in the left and right panels for elongated and pancake-like single-wells, respectively.} 
} 
\label{fig1_GS}
\end{figure}

Our analysis extends from (i) an elongated double-well characterized by $(\omega_x,\omega_y,\omega_z)= 2\pi \times (19,53,81) \rm Hz$ (with the single wells being similar to those in~\cite{Tanzi2019b,Bottcher2019}) to (ii) a pancake double-well with   $(\omega_x,\omega_y,\omega_z)= 2\pi \times (43,43,133) \rm Hz$ (as in~\cite{Klaus2022}). 
The lowest panels of Fig.~\ref{fig1_GS} show the contours of the potentials cut through $z=0$ for $D=10 \hbar \omega_x/l_{\rm osc}$. Also, the upper panels (a)-(c) and (d)-(f) represent the dBEC configurations in the left well before the quench for the elongated and planar single wells respectively. 
The characteristic timescales determined by the trap frequency accordingly correspond to $\omega_x^{-1}=8.4$~ms and $\omega_x^{-1}=3.7$~ms in the elongated and planar cases, 
with harmonic oscillator lengths  $l_{{\rm osc}}=\sqrt{\hbar/(m \omega_x)}=1.8{\rm \mu m}$  and $1.2{\rm \mu m}$, respectively. 
Another important timescale is set by the energy difference among the pre- and post-quench states, $\tau_{D_{{\rm in}}}=1/\Delta \mathcal{E}_{D_{\rm in}}$, where $\Delta \mathcal{E}_{D_{\rm in}}\equiv \mathcal{E}_{D_{{\rm in}}}- \mathcal{E}_{D_{\rm fi}}$. Here, $\mathcal{E}_{D_{{\rm in}}}$ and $\mathcal{E}_{D_{{\rm fi}}}$ refer to the energy of the dBEC at the initial ($D_{{\rm in}}=D$) and final ($D_{{\mathrm fi}}=0$) tilt strength. 
The interaction dependence of $\Delta \mathcal{E}_{D_{\rm in}}$ is weak for the considered values of $\epsilon_{{\rm dd}}$. 
We thus provide an average value for each $D_{{\rm in}}$. 
In this sense, relevant timescales obtained from this energy difference correspond to $\tau_{D_{{\rm in}}=10} \approx 40$ms and $\tau_{D_{{\rm in}}=7} \approx 20$ms in both settings. 

In what follows, we explore a $^{164}$Dy dBEC under the influence of the energy offset for specific values of $\epsilon_{{\rm dd}}$ corresponding to a SF, a SS, and a DL in the initial state. After preparing the initial configuration, we perform a quench of the tilt strength from a finite value to zero, and monitor the system dynamics as the energy offset vanishes for up to one second. 

\section{Initial states in the tilted double-well}\label{initialization}

Let us begin by investigating the ground-state phases of the dBEC as sketched in Fig.~\ref{fig1_GS}. 
For simplicity, we employ a tilt strength $D=7\hbar \omega_x/l_{{\rm osc}}$ providing an energy offset among the wells which ensures that the entire dBEC is initially trapped in the left part of the potential. 
To obtain the ground states, we employ the imaginary time propagation method in the eGPE of Eq.~(\ref{eGPE}) using the split-step Crank-Nicolson~\cite{CrankNicolson1947} approach (see also Appendix~\ref{numerics}). 
Characteristic iso-surfaces of the three-dimensional density, $n(x,y,z)$, are shown in Fig.~\ref{fig1_GS} for different values of $\epsilon_{{\rm dd}}$. 
Recall that the integrated  spatial density distribution (over tightly confined directions) showcasing crystal arrays can be experimentally probed by {\it in-situ} imaging~\cite{Sohmen2021,Norcia2021a}. 

\begin{figure} 
\centering
\includegraphics[width = 0.48\textwidth]{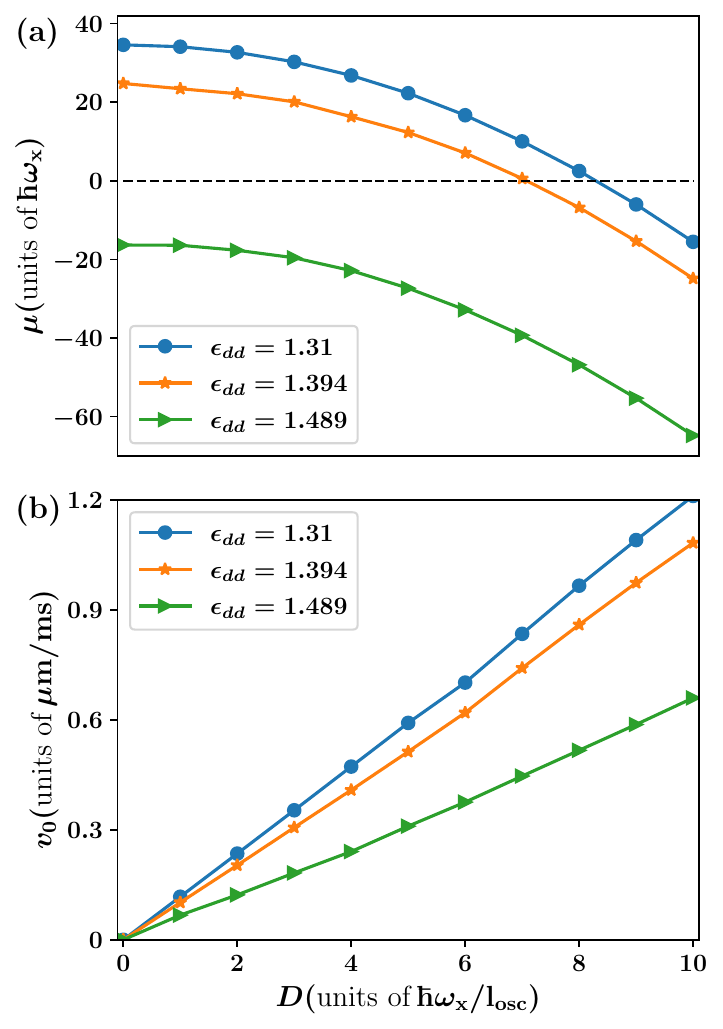}
\caption{(a) Dependence of the dBEC chemical potential in the elongated double-well for varying tilt strength, $D$, and specific relative interaction coefficients $\epsilon_{ {\rm dd}}$ (see legend). The interplay among $D$ and $\epsilon_{{\rm dd}}$ dictates the ground-state phases from SF to SS and droplet lattice. (b) The initial velocity, $v(0)\equiv v_0$, of the dBEC center-of-mass as a function of the tilt strength for several $\epsilon_{ {\rm dd}}$ (see legend). 
It features an interaction-dependent linear behavior with the SF ($\epsilon_{ {\rm dd}}=1.31$) possessing a larger velocity. 
The modification in the slope of $v_0$ around $D=6\hbar \omega_x/l_{{\rm osc}}$ occurs since above this value the entire dBEC is located at the left part of the double-well.} 
\label{fig1_chem}
\end{figure}

The structural configurations depend on both the magnitude of $\epsilon_{ {\rm dd}}$ and the confinement geometry. 
For relatively small values of $\epsilon_{{\rm dd}}$ such as $\epsilon_{{\rm dd}} \approx 1.31$  where the contact interaction dominates over the long-range anisotropic dipolar interaction, a non-modulated SF state emerges. It is characterized by the 
typical homogeneous density profile along the $x$-direction [Fig.~\ref{fig1_GS}(a)] for the elongated single well, and a pancake distribution in the $x$-$y$ plane [Fig.~\ref{fig1_GS}(d)] in the circular single well, respectively. 
The distributions are  
compressed across the tightly confined directions along $y$ and $z$ in the elongated trap and across the $z$-direction in the case of the planar trap, see Sec.~\ref{theoy_sec}. 
As expected, in both cases, the SF state has positive energy and chemical potential, here, for $D=7\hbar \omega_x/l_{{\rm osc}}$
(see also Fig.~\ref{fig1_chem}(a)). 

\begin{figure*}
\centering
\includegraphics[width = 1.0\textwidth]{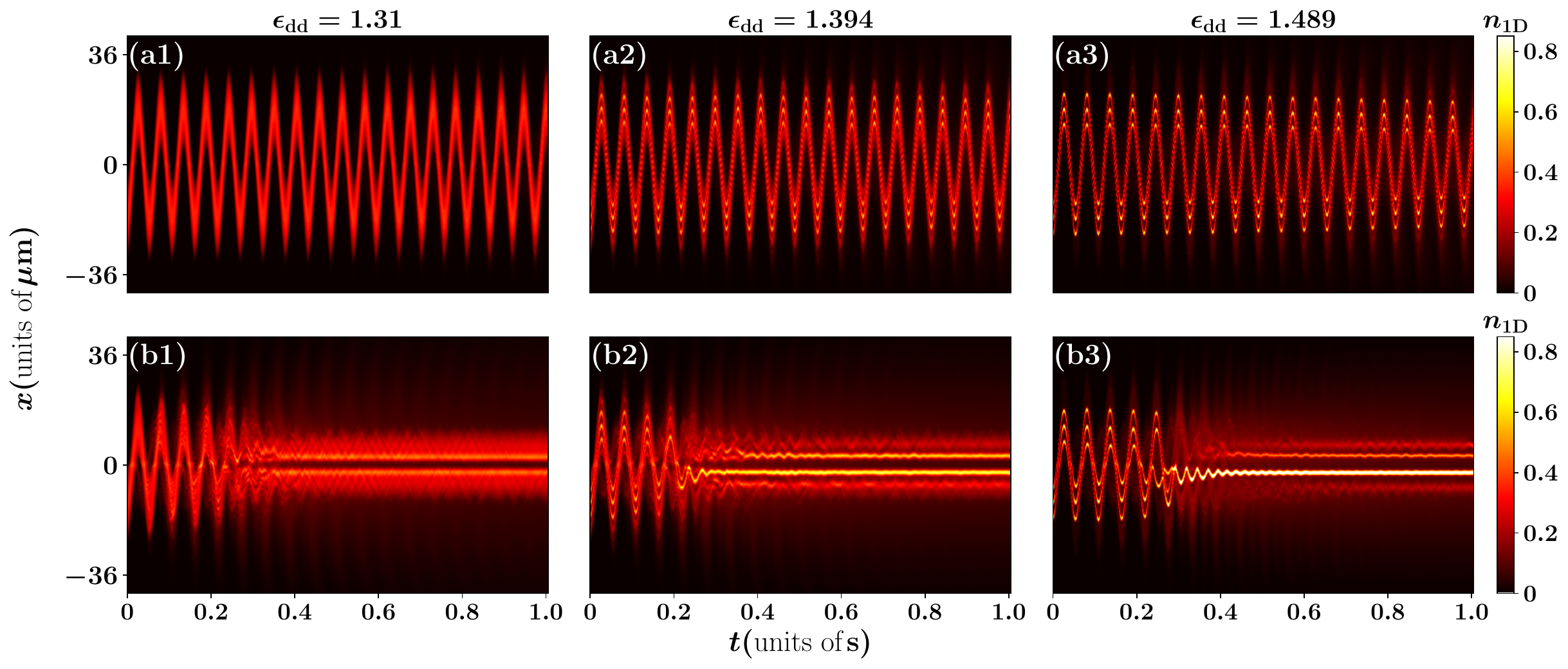}
\caption{Dynamics in the elongated double-well, triggered by quenching the tilt strength from (a1)-(a3)  $D=10\hbar \omega_x/l_{{\rm osc}}$ and (b1)-(b3) $D=7\hbar \omega_x/l_{{\rm osc}}$ to $D=0$. In the panels from left to right,
the system is in the SF with $\epsilon_{{\rm dd}}=1.31$, a SS characterized by $\epsilon_{{\rm dd}}=1.394$ and a droplet array with $\epsilon_{{\rm dd}}=1.489$, respectively. 
The colorbar represents the integrated density $n_{{\rm 1D}}(x,t)$ after the quench  in units of $N/l_{\rm osc}$, where $N=40000$ and $l_{\rm osc}=1.80 {\rm \mu m}$.
The tunneling frequency is nearly un-altered for different values of $\epsilon_{{\rm dd}}$, but the amplitude is reduced for 
increasing $\epsilon_{{\rm dd}}$. 
It is evident that the tunnelling properties of the dBEC depend strongly on its initial velocity as quantified by the initial $D$ 
value, {\textit {i.e.}}, for decreasing $D$ the dBEC density smudges over both wells for longer evolution times. }
\label{densities_1Dtunel}
\end{figure*}

Increasing $\epsilon_{{\rm dd}}$ while maintaining the confinement and dipolar direction eventually renders the system self-bound,   
characterized by a negative chemical potential. 
For instance, when $\epsilon_{{\rm dd}} \approx 1.4$ for the elongated  single-well (and similarly, for $\epsilon_{ {\rm dd}}=1.49$ in the case of the pancake-like double-well) we observe the formation of a periodic, density-modulated pattern in the left well\footnote{Concerning a state of the dBEC that is entirely trapped in one of the wells, the transition from the SF towards the SS and DL phases can also be seen in the momentum distribution where in contrast to the single-peak structure of the SF additional higher-lying ones accumulate in both the SS and droplet regimes.}.  
The individual density peaks\footnote{ The number of the generated density humps increases for larger atom number while keeping all other system parameters fixed as it was also discussed {\it e.g.} in Ref.~\cite{Poli2021}. These structures appear in the transverse direction as elongated filaments.} are phase-coherently inter-linked by a coherent background SF and a SS state forms. 
For larger $\epsilon_{ {\rm dd}}$, the dipolar interactions become dominant. 
This leads to a vanishing SF background and a simultaneous stronger spatial localization of the aforementioned density humps, see Fig.~\ref{fig1_GS}(c), (f). 
In this droplet phase, the self-bound character is related to exhibiting large negative energy or equivalently chemical potential [Fig.~\ref{fig1_chem}(a)]. 
Figure~\ref{fig1_GS}(c) depicts a characteristic DL density distribution in an elongated geometry. However, in the planar geometry, 
instead, a central droplet is surrounded by a ring-shaped SF configuration, see Fig.\ref{fig1_GS}(f), as discussed in~\cite{Bisset2016}.

Increasing the energy offset favors the DL formation and can lead to a negative chemical  potential [Fig.~\ref{fig1_chem}(a)].
The offset thus also enables a crossover to  self-bound states in the double-well. 
For a smaller energy offset between the wells a certain fraction of population can still reside in the right part of the double-well. 
The interaction-dependent critical tilt value, where the entire cloud is solely trapped in the left well, is lower for larger $\epsilon_{ {\rm dd}}$, \textit{e.g.} $D<4\hbar \omega_x/l_{{\rm osc}}$ for $\epsilon_{ {\rm dd}}=1.49$, and increases for reduced $\epsilon_{{\rm dd}}$, \textit{e.g.} $D<6\hbar \omega_x/l_{{\rm osc}}$ for $\epsilon_{{\rm dd}}=1.36$.
Furthermore, through the adjustment of $D$, it is possible to prepare more intricate density patterns arising from populations located in both wells. For instance, when $D=2\hbar \omega_x/l_{{\rm osc}}$, two droplets  form in the left well, while a single-slanted DL structure appears in the right well. Such configurations, influenced by interference and restricted initial velocities dictated by the value of $D$, have a significant impact on the resulting dynamical behavior, as it will be discussed in the following. 

\begin{figure}
\centering
\includegraphics[width =0.44\textwidth]{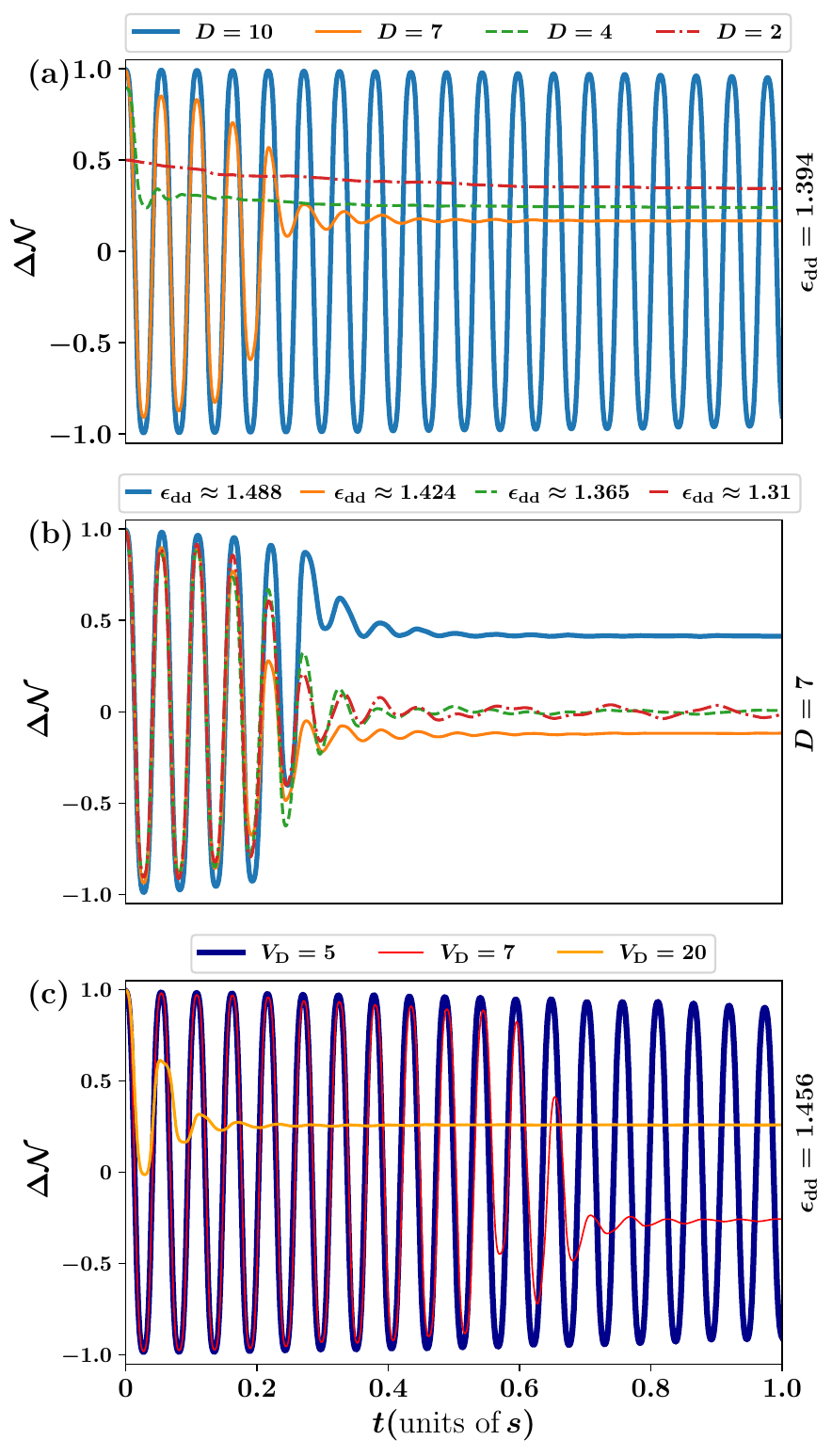}
\caption{Time-evolution of the population imbalance, $\Delta \mathcal{N}(t)$, between the left and right wells for (a) different initial tilts $D$ (see legend) at fixed $\epsilon_{{\rm dd}} =1.394$ referring to a SS and (b) distinct relative interactions, $\epsilon_{{\rm dd}}$ (see legend) when  $D=7\hbar \omega_x/l_{{\rm osc}}$. It is evident that a decreasing tilt strength slows down tunneling, whilst for a constant $D$ population imbalance at long evolution times is suppressed only within the SF phase and becomes maximal for a droplet array. 
(c) $\Delta \mathcal{N}(t)$ in the SS case with $\epsilon_{{\rm dd}} =1.456$ and using an initial $D=7\hbar \omega_x/l_{{\rm osc}}$ for several barrier heights $V_D$ (see legend). 
An increasing $V_D$ leads to a macroscopic self-trapping regime.
The dBEC consists of $N=40000$ $^{164}$Dy atoms confined in a quasi-1D double-well. The tilt strength $D$ of the double-well is measured in units of $\hbar \omega_x/l_{{\rm osc}}$.} \label{imbalance_1D}
\end{figure}

\section{Dynamics in an elongated double-well}\label{tunnel1D}

To explore the dynamical properties of the dBEC in a double-well, we first consider an elongated geometry along the x-direction.  
The system is quenched by instantaneously (at $t=0$) switching-off the external tilt, from a value $D>0$ to $D=0$. For the impact of a time-dependent ramp-down of the tilt on the tunneling dynamics and the persistence of the discussed tunneling regimes for different ramp-rates see Appendix~\ref{td_quench}. 
Naturally, the strength of the tilt potential is related to the initial velocity imparted on the dBEC directly after performing the $D=0$ quench. 
Namely, a larger initial $D$ refers to an increasing velocity. 
This can be readily verified by determining the velocity of the center-of-mass of the dBEC, $v(t=0)\equiv v_0$, provided in Fig.~\ref{fig1_chem}(b) for various interactions as a function of $D$.  The velocity, herein, is defined as the time-derivative of the center-of-mass coordinate $X_{{\rm CM}}(t)= \int_{-x_0}^{x_0}  \mathrm{dx}~x n_{{\rm 1D}}(x,t)$, where $n_{ {\rm 1D}}(x,t)=\int \mathrm{dydz}~n(x,y,z,t)$ denotes the one-dimensional (1D) integrated density and $\pm x_0/2$ is the location of the employed hard-wall boundaries along the $x$-direction. These are, of course, chosen sufficiently wide such that they do not affect our results.   
The dependence of $v_0$ on the interaction strength is clearly seen. For larger $\epsilon_{{\rm dd}}$ values and a fixed $D$, \textit{i.e.} towards the droplet regime, the velocity decreases, reflecting the rigidity of the state.
%A central aim of our investigation is to reveal the interplay of the dBEC interactions ($\epsilon_{{\rm dd}}$) and initial velocity (dictated by $D$) on the emergent tunneling properties. 

To visualize the overall dynamical response of the elongated dBEC we invoke the underlying integrated density, $n_{ { \rm 1D}}(x,t)$ [Fig.~\ref{densities_1Dtunel}]. Moreover, in order to determine the emergent tunneling behavior, we monitor the population imbalance~\cite{Mistakidis2022} among the left (L) and right (R) wells,
\begin{equation}
\begin{split}
\Delta \mathcal{N} (t)= \frac{1}{N} \bigg[\int_{-x_0/2}^0  \mathrm{dx}~n_{ {\rm 1D}}(x,t) &- \int_{0}^{x_0/2}  \mathrm{dx}~n_{ { \rm 1D}}(x,t) \bigg] \\& \equiv n_L(t)-n_{R}(t).  \label{prob_imbal} 
\end{split}
\end{equation} 
It allows to estimate both the frequency and amplitude of the induced macroscopic tunneling properties. 

As a starting point, we leverage a relatively large tilt strength, $D \geq 10\hbar \omega_x/l_{{\rm osc}}$, rendering the impact of the initial velocity $v_0$ (which is enhanced here, Fig.~\ref{fig1_chem}(b)) irrelevant. 
Indeed, for $D \geq 10\hbar \omega_x/l_{{\rm osc}}$, the entire dipolar gas is localized in the left double-well rendering the quench-induced tunneling process independent of the tilt strength. 
This enables us to unravel the ensuing interaction effects on the dynamics. 
Ramping-down $D$ leads to coherent oscillations of the dBEC with a frequency that is weakly dependent on the relative interactions $\epsilon_{{\rm dd}}$, see Fig.~\ref{densities_1Dtunel}(a1)-(a3) and Fig.~\ref{imbalance_1D}(b).  
Notice, however, that the oscillation amplitude is reduced for increasing $\epsilon_{{\rm dd}}$, see also the discussion below and Fig.~\ref{imbalance_1D}(b). 
In the case of a SF ($\epsilon_{{\rm dd}} \approx 1.31$) the entire dBEC coherently moves between the left and right wells with constant amplitude, and period\footnote{Notice that $T_{{\rm Rabi}}\sim  2 \pi/(\omega_x)$ for large $D$, meaning that the tunneling is essentially a collective center-of-mass oscillation, i.e. the dipole mode. However, a more accurate estimation of the tunneling period for all $D$ examined herein would require knowledge of the ensuing energy gap among the wells. In this context, a larger gap enforced by increasing tilt strength leads to a smaller period.} $T_{{\rm Rabi}} \approx 55\rm$, in analogy to what is known for regular BECs~\cite{Smerzi1997,Albiez2005}. 
This response will be dubbed macroscopic collective tunneling regime. 

A similar dynamical behavior is also observed for a SS [Fig.~\ref{densities_1Dtunel}(a2) for $\epsilon_{{\rm dd}}=1.397$] and a droplet array [Fig.~\ref{densities_1Dtunel}(a3) for $\epsilon_{{\rm dd}}=1.49$]. 
Initially being localized in the left well, after quenching, the SS or DL crystal structures move towards the right side, and during this process come into close proximity with each other, particularly in the vicinity of the potential barrier 
(see Figs.~\ref{densities_1Dtunel} (a2) and (a3). Subsequently, the crystals move into the right well, and as they reach the right well's edge, they separate from each other again. This process repeats periodically. The maximum separation between the single crystals at the well's edge is more pronounced for larger $\epsilon_{{\rm dd}}$ due to the dominant dipolar interactions, which enhance repulsion among the density humps in the transverse plane [see Figs.~\ref{densities_1Dtunel}(a3) and (b3)].

\begin{figure}
\centering
\includegraphics[width =0.48\textwidth]{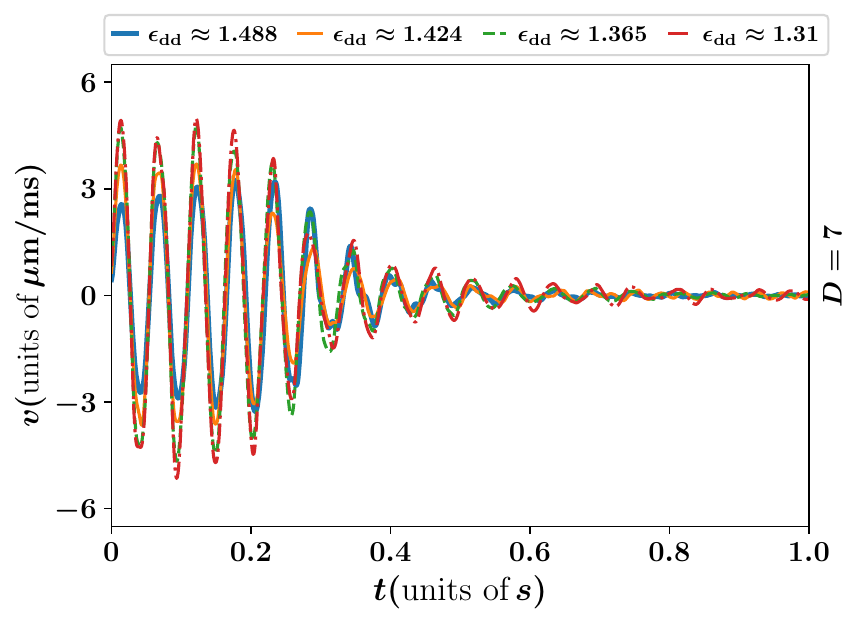}
\caption{Dynamics of the velocity, $v(t)$, of the dBEC center-of-mass coordinate for various $\epsilon_{{\rm dd}}$ (see legend) keeping $D=7\hbar \omega_x/l_{{\rm osc}}$. 
Apparently, upon considering fixed $D$ and increasing $\epsilon_{{\rm dd}}$ towards the droplet regime leads to a progressively smaller velocity signifying the crystalline nature of the dBEC. 
At long evolution times the dBEC relaxes and its velocity vanishes. The tilt strength $D$ of the double-well is expressed in terms of $\hbar \omega_x/l_{{\rm osc}}$.} \label{velocity_1D}
\end{figure}

\begin{figure*}
\centering
\includegraphics[width = 1.0\textwidth]{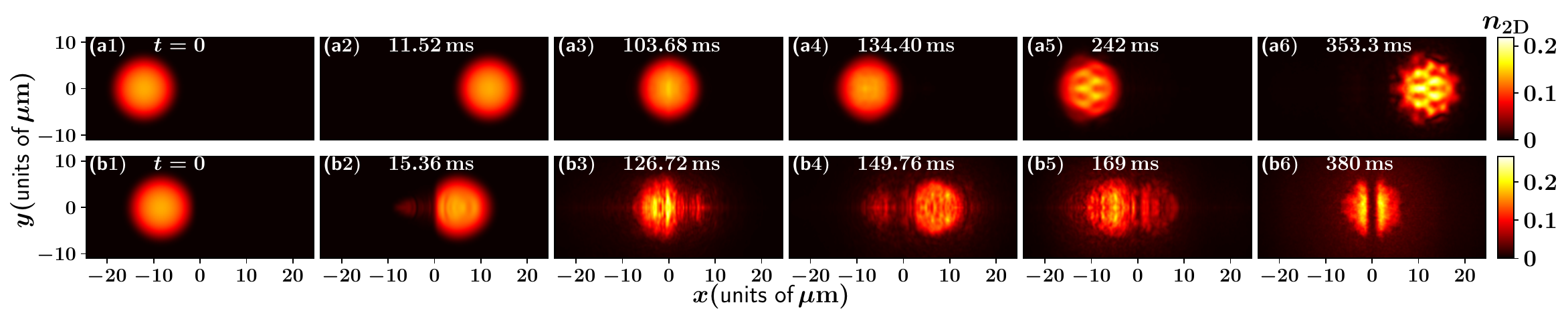}
\caption{Snapshots of the integrated 2D density $n_{\rm 2D}(x,y,t)$ visualizing the tunneling of a SF configuration for $\epsilon_{{\rm dd}}=1.31$ when the initial tilt strength of the double-well is set to (a1)-(a6) $D=10\hbar \omega_x/l_{{\rm osc}}$ and (b1)-(b6) $D=7\hbar \omega_x/l_{{\rm osc}}$. 
The tunneling behavior depends on the original $D$ value. 
For $D=10\hbar \omega_x/l_{{\rm osc}}$, \textit {i.e.}, large initial velocities, it leads to persistent oscillations of the SF among the wells and its motional excitation at longer evolution times. 
However, a reduced velocity (here quantified by $D=7\hbar \omega_x/l_{{\rm osc}}$) results in a progressive damping of the tunneling and its subsequent locking. 
The evolution is induced by suddenly suppressing the tilt strength. 
The colorbar signifies $n_{ { \rm 2D}}(x,y,t)$ in units of $N/l_{\rm osc}$, where $N=40000$ and $l_{\rm osc}=1.2 {\rm \mu m}$.} \label{densities_2D_dyn_tunel}
\end{figure*}

A reduction of the tilt strength (or equivalently the initial velocity, Fig.~\ref{fig1_chem}(b)) leads to significant alterations of the dynamical behavior. Comparing  Fig.~\ref{densities_1Dtunel}(a1)-(a3) with Fig.~\ref{densities_1Dtunel}(b1)-(b3) reveals a different response, displaying intricate tunneling dynamics. 
At short evolution times, $t\approx \tau_{D_{{\rm in}}=7}$, the dBEC oscillates back and forth between the wells. 
However, owing to the reduced velocity at $t=0$, the dBEC gradually starts to be partially transmitted to the other well while a relatively small fraction is reflected back. 
As more particles are reflected in the course of the evolution, individual density humps become less prominent both in the SS [Fig.~\ref{densities_1Dtunel}(b2)] and the DL [Fig.~\ref{densities_1Dtunel}(b3)].  
Eventually, after four oscillation periods ($t >> \tau_{D_{{\rm in}=7}}$) a certain population remains trapped within each well and tunneling locks. 
Indeed, employing $D=7\hbar \omega_x/l_{{\rm osc}}$ which still ensures the confinement of the dBEC within the left well at $t=0$, 
we observe that the smaller initial velocity (as compared to the case with $D=10\hbar \omega_x/l_{{\rm osc}}$) results in a gradual suppression  of the inter-well tunneling, see also $\Delta \mathcal{N}(t)$ in Fig.~\ref{imbalance_1D}(a). 
A further decrease of $D$ favors an appreciable part of the population to initially also reside in the right well. This eventually leads to suppression of the tunneling due to the relatively weak exerted momentum by the quench, see the values of  $\Delta \mathcal{N}(t)$ for $D=2\hbar \omega_x/l_{{\rm osc}}$ for example in Fig.~\ref{imbalance_1D}(a). This resembles a type of self-trapping, as it has been also realized in regular BECs~\cite{Milburn1997,Albiez2005,Levy2007,Gati2007}. 

For larger $\epsilon_{\rm dd}$, different droplet lattice configurations can be realized in the left well compared to the right well. In this case, the individual crystals or SSs undergo multifrequency intrawell oscillations of varying amplitude induced by the tunneling of the background SF, see also the explicit time-evolution in~\cite{video_data}. 
We note here that removing the small offset between the wells is essentially equivalent to imparting a small kick velocity to the crystal structures. The  vibrational patterns emanate from the activation of the different underlying normal modes, as demonstrated in~\cite{Mukherjee2023}. The main features of the above-described tunneling behavior remain robust (at least up to certain times) even in the presence of three-body losses, which are customarily present in experiments, as showcased in Appendix~\ref{3body}. However, in this case the tunneling is suppressed for longer evolution times and e.g. the SS or DL character is lost due to atom losses. 

A smudging of density occurs for the SF, SS and DL regimes as can be readily seen by inspecting the population imbalance, $\Delta \mathcal{N}(t)$, shown in Fig.~\ref{imbalance_1D}(b) for varying $\epsilon_{{\rm dd}}$ and fixed $D=7\hbar \omega_x/l_{{\rm osc}}$. 
In particular, we deduce that the condensate density smudges at a faster rate for a SS ($\epsilon_{{\rm dd}}=1.424$) as compared to the other phases, an outcome that is attributed to the enhanced interference of the individual crystals due to the background SF\footnote{The same observations in terms of the tunneling regimes can be made by inspecting the dynamics of the integrated current $J(t)=\frac{\hbar}{2mi} \int \mathrm{dx dy dz} \big(\Psi^* \frac{\partial \Psi}{\partial x}- \Psi \frac{\partial \Psi^*}{\partial x}\big)$, not shown for brevity.}. 
Moreover, the population imbalance in the long time-evolution after the termination of the tunneling is found to depend on the dBEC phase and thus, the interactions. 
With increasing $\epsilon_{dd}$, atoms exhibit a stronger tendency to self-bind, leading to their preferential accumulation within a specific well in the long time-evolution and causing a more pronounced atom imbalance. A larger dilute SF component, occurring for weaker $\epsilon_{\rm dd}$,  however, facilitates the progressive restoration of inter-well population balance.    
As such, $\Delta \mathcal{N}(t)$ in the course of the tunneling may serve as a probe to identify the crystalline nature of the dBEC.
The suppression of the tunneling is also supported by the behavior of the velocity of the center-of-mass showcased in Fig.~\ref{velocity_1D}. Keeping the initial tilt strength fixed, for example at $D=7\hbar \omega_x/l_{{\rm osc}}$, the velocity in the course of the dBEC evolution is overall reduced towards the self-bound state regime realized for larger $\epsilon_{{\rm dd}}$. 
This can be traced back to the enhanced rigidity of the droplet arrays. 

Naturally, the barrier height has a crucial impact on the tunneling, as is also known from short-range interacting BECs~\cite{Gati2007,Mistakidis2022,Keshavamurthy2011}. 
To explicate this dependence we consider a SS state at $\epsilon_{{\rm dd}}=1.456$ experiencing a tilt of $D=7\hbar \omega_x/l_{{\rm osc}}$ and measure $\Delta \mathcal{N}(t)$ for various $V_D$ presented in Fig.~\ref{imbalance_1D}(c). 
It becomes apparent that by increasing $V_D$, which equivalently means that the involved energy gaps become larger~\cite{Gati2007}, it is possible to transition to  distinct tunneling regions. 
In other words, it is possible to adjust the dBEC response from coherent density oscillations, as here for $V_D=5\hbar \omega_x$, to density smudging and tunneling locking at long times for instance when $V_D=7\hbar \omega_x$, and  vanishing inter-well tunneling at $V>20\hbar \omega_x$. 

\section{Dynamics in a pancake double-well }\label{tunnel2D}

Let us now investigate how the above-described phenomena depend on the dimensionality of the system, and turn to the pancake-like double-well. The dBEC is initialized in the tilted double-well assembling in different configurations according to the value of $\epsilon_{{\rm dd}}$ as depicted in Fig.~\ref{fig1_GS}(d)-(f). As before, 
the time-evolution is induced via quenching the tilt strength $D$ from a finite value down to $D=0$.
Similar to the elongated geometry [Fig.~\ref{fig1_chem}(b)], the initial velocity, $v_0$, exhibits a linear increase for larger $D$ and fixed $\epsilon_{{\rm dd}}$, while being reduced for larger $\epsilon_{{\rm dd}}$ and constant $D$ (not shown). 
The overall dynamical behavior is characterized through the two-dimensional (2D) integrated density defined as $n_{ { \rm 2D}}(x,y,t)=\int \mathrm{dz}~n(x,y,z,t)$ and the respective population imbalance $\Delta \mathcal{N}(t)=(1/N)[\int_{-x_0/2}^0\int_{-y_0}^{y_{0}}\mathrm{dxdy}~n_{{\rm{ 2D}}}(x,y,t)-\int_{0}^{x_0/2}\int_{-y_{0}}^{y_0}\mathrm{dxdy}~ n_{{\rm{ 2D}}}(x,y,t)]$, with $\pm x_0/2$ [$\pm y_0/2$] denoting the position of the hard-walls in the $x$ [$y$] direction.

\begin{figure*}
\centering
\includegraphics[width = 1.0\textwidth]{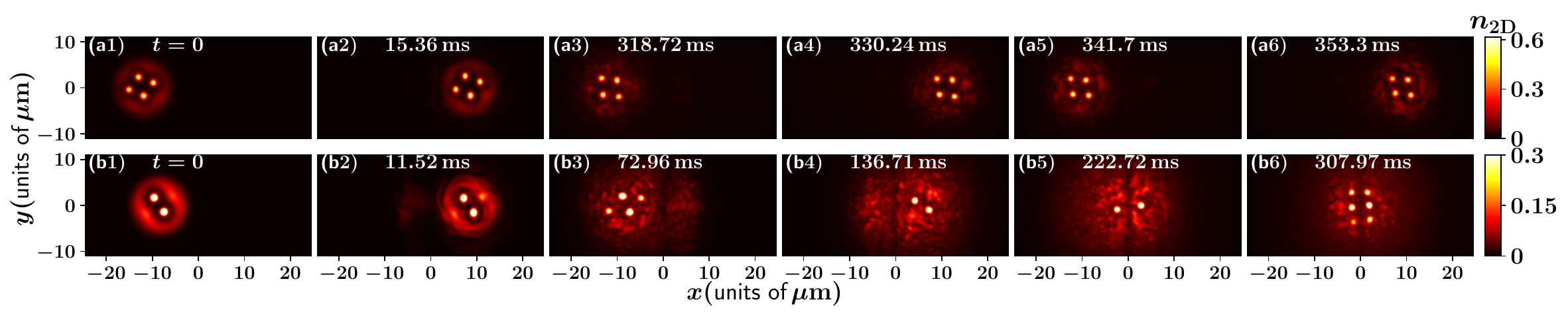}
\caption{Density snapshots demonstrating the tunneling behavior of a SS for $\epsilon_{{\rm dd}} \approx 1.49$ using an initial inter-well energy offset characterized by (a1)-(a6) $D=10\hbar \omega_x/l_{{\rm osc}}$ and (b1)-(b6) $D=7\hbar \omega_x/l_{{\rm osc}}$. 
In the case of $D=10\hbar \omega_x/l_{{\rm osc}}$, the SS oscillates back and forth maintaining its shape with the involved crystals featuring quadrupole excitations. 
In contrast, for $D=7\hbar \omega_x/l_{{\rm osc}}$ dynamically unstable crystals appear in the course of the evolution and eventually at long-times tunneling locks. 
The colorbar represents $n_{ { \rm 2D}}(x,y,t)$ in terms of $N/l_{\rm osc}$, with $N=40000$ and $l_{\rm osc}=1.2 {\rm \mu m}$.} \label{densities_2DSS}
\end{figure*}

\begin{figure}
\centering
\includegraphics[width = 0.45\textwidth]{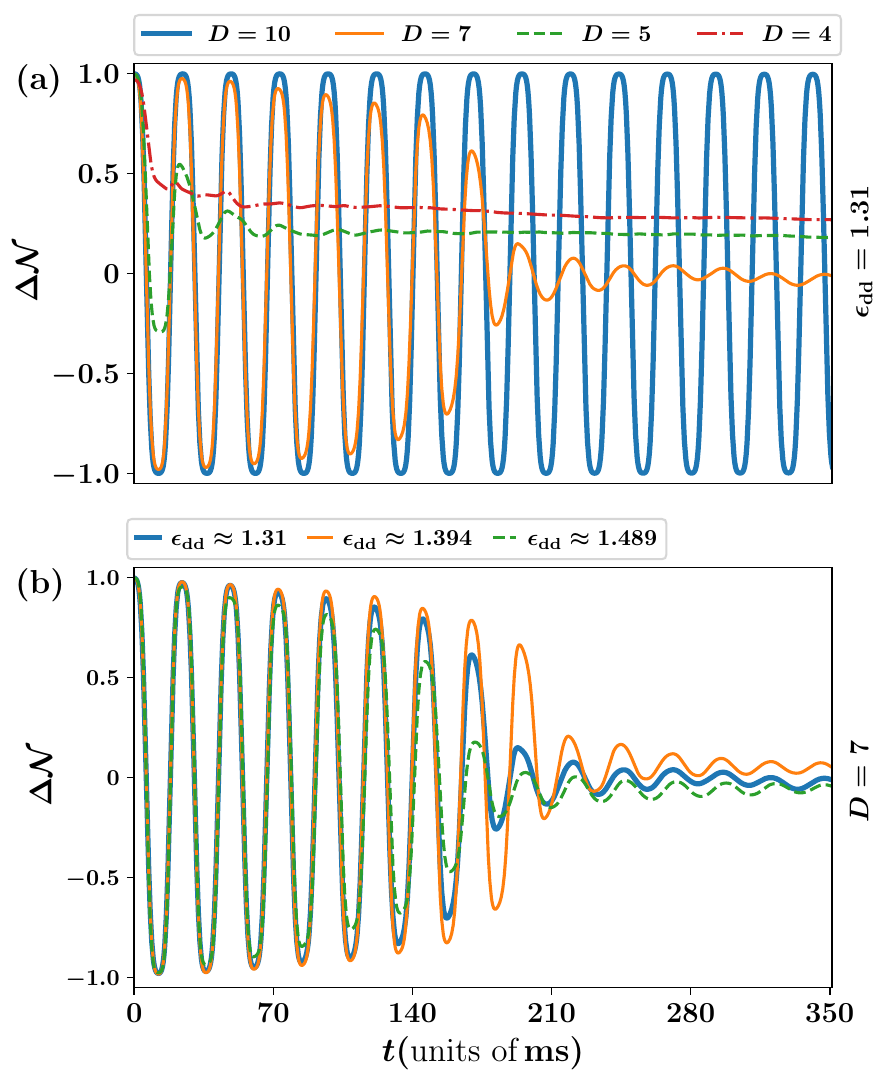}
\caption{Dynamics of the inter-well population imbalance, $\Delta \mathcal{N}(t)$, in the quasi-2D double-well when considering (a) various tilt strengths $D$ (see legend) and $\epsilon_{{\rm dd}} =1.31$ corresponding to a SF or (b) using different $\epsilon_{{\rm dd}}$ (see legend) and $D=7\hbar \omega_x/l_{{\rm osc}}$. A reduced energy offset results to a gradual suppression of tunneling. 
Moreover, $\Delta \mathcal{N}(t)$ in the region of $D$ where tunneling locks vanishes within the SF phase but it remains finite in the SS. 
The evolution of the quasi-2D dBEC with $N=40000$ $^{164}$Dy atoms is triggered by quenching the initial potential offset $D$ to zero.} \label{populations_2D_dyn}
\end{figure}

Considering a large energy offset  at $t=0$, quantified by $D=10\hbar \omega_x/l_{{\rm osc}}$, we find that irrespective of $\epsilon_{{\rm dd}}$ the entire dBEC undergoes regular oscillations between the left and right wells with constant amplitude and frequency throughout the time evolution. 
This behavior can be directly seen in the densities of a SF [SS] depicted in  Fig.~\ref{densities_2D_dyn_tunel}(a1)-(a6) [Fig.~\ref{densities_2DSS}(a1)-(a6)] but also in the  corresponding inter-well population imbalance provided in Fig.~\ref{populations_2D_dyn}(a) exemplary for the SF case. 
Focusing on the evolution of the SF state with $\epsilon_{{\rm dd}}=1.31$, there are two main distinct dynamical stages:  
At early evolution times, the SF  coherently oscillates back and forth characterized by a Rabi frequency $T_{{\rm Rabi}} \approx 24 \rm ms \approx 2 \pi / \omega_x$ and without experiencing any noticeable density deformation. While later on [Fig.~\ref{densities_2D_dyn_tunel}(a5), (a6)], the density distortion gradually enhances, likely due to the trigger of underlying collective modes~\cite{Ronen2006,Guo2019,Natale2019,Tanzi2019b} after multiple collisions of the condensate with the central barrier.  

The tunneling behavior of a SS configuration obtained at $\epsilon_{dd}\approx 1.49$ with fixed $D=10\hbar \omega_x/l_{{\rm osc}}$ is demonstrated in Fig.~\ref{densities_2DSS}(a1)-(a6). 
Interestingly,  we observe the persistent inter-well  oscillatory motion as well as angular oscillation of the entire SS. However, the original SS retains its shape throughout the time evolution with the background superfluid  exhibiting spatial undulations [Fig.~\ref{densities_2DSS}(a3), (a4)] stemming in part from the particle flow between the constituting crystals and in part from the collision with the central barrier. 
The above response substantiates the rigidity of the droplet crystal arrangement, with the distance of the crystals remaining almost intact, and the dilute superfluid nature of the background. 
Indeed, the involved droplet crystals share a relatively small density overlap leading to debilitated particle flow among the droplets. 
As a consequence, the atom number within each droplet is not appreciably modified in the time evolution. 

As in the elongated geometry, a reduction of the initial tilt strength leads to significant modifications of the tunneling behavior due to the comparatively smaller initial velocity. 
Characteristic instantaneous density profiles of a SF dBEC ($\epsilon_{{\rm dd}}=1.31$) following a quench from $D=7\hbar \omega_x/l_{{\rm osc}}$ to $D=0$ are provided in Fig.~\ref{densities_2D_dyn_tunel}(b1)-(b6). 
It can be readily seen that the SF tunnels from the left to the right well [Fig.~\ref{densities_2D_dyn_tunel}(b1), (b2)] but the initial velocity is not sufficient for the entire cloud to be fully transmitted. 
As such, there is  a fraction of reflected density remaining back and becoming more pronounced during the dynamics; see, for example, Fig.~\ref{densities_2D_dyn_tunel}(b3)-(b5), which is characterized by enhanced spatial delocalization. 
This progressive collocation of density in both wells results in a vanishing population imbalance and tunneling ``locks" as visualized in the density snapshot of Fig.~\ref{densities_2D_dyn_tunel}(b6) and explicitly captured via $\Delta \mathcal{N}(t)$ in Fig.~\ref{populations_2D_dyn}(a). 
We remark that using smaller initial energy offsets, \textit{viz.}, $D \leq 5\hbar \omega_x/l_{{\rm osc}}$ the distribution of the dipoles takes place in both left and right wells. 
Following a quench to $D=0$ leads to suppression of tunneling as can be deduced from $\Delta \mathcal{N}(t)$, illustrated in Fig.~\ref{populations_2D_dyn}(a)  for $D=5\hbar \omega_x/l_{{\rm osc}}$ and $D=4\hbar \omega_x/l_{{\rm osc}}$, and eventually to solely intrawell dynamics of the underlying density fractions; see also in~\cite{video_2D_smalloffset}. 

Similar effects occur even when a SS is considered with smaller tilt. In Fig.~\ref{densities_2DSS}(b1)-(b6), we demonstrate 2D densities at different time-instants for $\epsilon_{{\rm dd}}=1.49$.  
For $D=7\hbar \omega_x/l_{{\rm osc}}$, the initial SS possesses only two-humps [Fig.~\ref{densities_2DSS}(b1)]. When performing the $D=0$ quench, the crystals, at early-time dynamics, oscillate together back and forth between the two wells, keeping their distance almost fixed. 
Simultaneously, a fraction of atoms is reflected from the barrier. Such reflection and transmission phenomena give rise to unequal populations in both wells during the dynamics. 
As a consequence, there are various unstable droplet configurations, such as the rhombic one depicted in Fig.~\ref{densities_2DSS}(b3), or two droplet configurations [Fig.~\ref{densities_2DSS}(b4)-(b5)], appearing during the dynamics. 
In the long time ($t > 310$ms) dynamics, atoms gradually accumulate in both wells in a symmetric manner, and 
a three-droplet configuration in each well can be seen on either side [Fig.~\ref{densities_2DSS}(b6)]. Consecutively, tunneling locks with a small inter-well population imbalance as can be seen in Fig.~\ref{populations_2D_dyn}(b). 
It is important to mention that the robustness of the SS structure during the tunneling dynamics depends strongly on the presence of the background SF which connects the individual crystals. Indeed, a more pronounced SF background eases particle transfer between the crystals, thus rendering them more susceptible to distortions and making them less rigid.
 
\section{Conclusions $\&$ perspectives}\label{conclusion} 

We investigated the ground state and  the tunneling dynamics of a dBEC trapped in an initially tilted double-well potential encompassing both elongated and pancake trap geometries. 
To model these systems and their dynamical behavior, we employed the extended Gross-Pitaevskii framework that incorporates first-order quantum corrections. 
The different phases are initialized by tuning the relative strength ratio between dipolar and short-range interactions while maintaining the tilt strength. 
The  inter-well energy offset imposed by the tilt strength effectively modifies the trapping volume of the dBEC. Thus, even for fixed interactions, when changing the tilt strengths  a transition from a SF to SS and DL state may occur, decreasing the chemical potential to negative values. 
The tilt value, furthermore, governs the initial center-of-mass velocity of the dBEC. For larger offsets and fixed interactions, the velocity monotonically increases. Conversely, for a specific tilt and varying interactions, the velocity decreases as the system transitions from the SF phase towards the SS and DL phases, evincing the rigidity of the latter. 

The time-evolution of the system is initiated by suddenly releasing the initial inter-well energy offset, promoting the tunneling motion of the dBEC. For the elongated double-well, it is observed that for sufficiently large tilt strengths, the dBEC, regardless of being in the SF, SS, or DL  phase, undergoes collective oscillations with a definite amplitude and frequency. 
On the other hand, a smaller initial energy offset leads to the partial transmission and reflection of the dipoles during the dynamics. The resultant interference of these accumulated density fractions gives rise to a smudging of the density, evident in the intra-well population, and eventually tunneling locks. In this dynamical stage, the inter-well population imbalance is highly asymmetric for the droplet array, while it vanishes in the SF case. This characteristic provides a valuable means to investigate the nature of the individual phases. 
The rigidity of the crystal-like SS and droplet configurations is further supported by their smaller center-of-mass velocity compared to the SF phase during the dynamics.
Maintaining fixed interactions and tilt strength but increasing the barrier height, the dipoles consequently remain confined within the initial well. 

A similar dynamical response in terms of the aforementioned system parameters takes also place in quasi-2D but there are characteristic properties that are inherently related to the  dimensionality of SS and DL. 
Indeed, for sufficiently large tilt strengths the dBEC undergoes regular oscillations back and forth among the wells. It also features distinctive dynamical patterns depending on the interactions. 
A SS with suppressed background SF tunnels as a whole, maintaining its initial configurations which substantiates its rigidity.  This motion is accompanied by small-amplitude angular oscillations triggered by the quench. 
Exploiting a reduced energy offset leads to partial transmission events of the dipolar cloud and ultimately to tunneling locking caused by collocation of density in each well during the evolution. 
This dynamical regime is characterized by vanishing inter-well population imbalance.

There is a variety of possible extensions of our results in future endeavors. An imperative prospect is to testify the validity of the eGPE to adequately describe all the emergent tunneling channels or even examine geometries with a few lattice sites~\cite{wilsmann2018control,grun2022integrable}. 
Along these lines, it is worth to unravel the participation of possible  interband tunneling processes especially in the few- to many-body crossover relying, for instance, on ab-initio methods~\cite{Cao2017,Haldar2019}. 
In this context, it would be possible to achieve an accurate characterization of the induced  tunneling pathways in terms of the relevant energy gaps and possibly reveal higher-order, e.g. inter-band, tunneling channels. 
Additionally, the inclusion of temperature and its impact on the discussed tunneling properties is worth pursuing.   
A first step towards this direction could be to utilize a stochastic Gross-Pitaevskii model as it was done in~\cite{Bland2022a}. Exploiting dynamical frustration events, \textit{e.g.}, by ramping a lattice potential in SS and DL~\cite{Halperin2023}, to design exotic tunneling processes is another intriguing direction. The spontaneous generation of nonlinear wave structures stemming from the counterflow dynamics of two initially separated dBEC clouds is certainly desirable. 
Furthermore, a detailed study of the excitation of corresponding surface modes in dBECs (for example using parametric driving as it was achieved in regular  BECs~\cite{Engels2007,Kwon2021}) constitutes another interesting future prospect.  

\section*{Acknowledgements} 
S. I. M. and H. R. S. acknowledge support from the NSF through a grant for ITAMP at Harvard University. 
K. M. and S. M. R. are financially supported by the Knut and Alice Wallenberg Foundation (KAW 2018.0217) and  the  Swedish  Research  Council (2022-03654-VR).

\appendix

\begin{figure*}
\centering
\includegraphics[width = 1.0\textwidth]{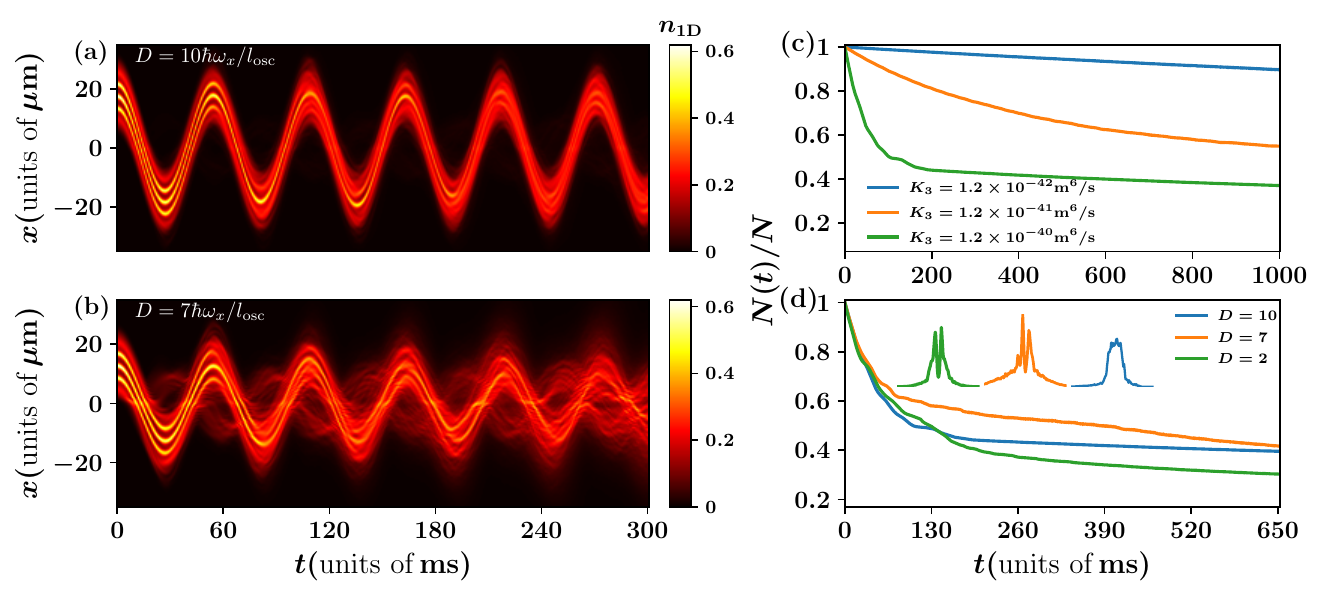}
\caption{Time-evolution of the integrated density, $n_{\rm 1D}(x,t)$, showcasing the tunneling dynamics in the presence of three-body losses with recombination rate coefficient $K_3 = 1.2 \times 10^{-40} \rm m^6/s$ for initial tilt strengths (a) $D=10\hbar \omega_x/l_{{\rm osc}}$ and (b) $D=7\hbar \omega_x/l_{ { \rm osc}}$. 
The tunneling processes of the $K_3=0$ case persist for early times and afterwards the SS character disappears due to prominent atom losses. 
The colorbar represents the integrated density $n_{{\rm 1D}}(x,t)$ in units of $N/l_{{\rm osc}}$, where $N=40000$ and $l_{\rm osc}=1.80 {\rm \mu m}$. 
Dynamics of the normalized atom number $N(t)/N(t=0)$ for (c) different recombination rate coefficients $K_3$ (see legend), focusing on $D=10 \hbar \omega_x/l_{ { \rm osc}}$ and (d) varying tilt strength $D$ (see legend). 
As expected, a more prominent reduction of $N(t)$ takes places for larger $K_3$. 
Inset of (d) depicts density profiles, $n_{\rm 1D}(x,t)$, at $t=500$ms and distinct $D$ values evincing the non-trivial effect of the tilt strength on the density localization.  
Apparently, the atom loss is crucially affected from the density localization, \textit{viz.}, it is enhanced for larger peak densities. 
In all cases, the relative interaction parameter $\epsilon_{\rm dd} = 1.394$.} \label{densities_1Dtunel_loss}
\end{figure*}

\section{Impact of three-body recombination in the tunneling dynamics}\label{3body}

The self-bound SS and droplet configurations are known to feature appreciable three-body losses~\cite{Maier2015,Chomaz2018} whose presence naturally destructs, in the experiment, the long-time observation and characteristics of these structures. 
For this reason, in what follows, we aim to expose the effect of the three-body loss rate in the tunneling properties \textit{e.g.}  of the elongated SS at $\epsilon_{{\rm dd}}=1.394$. 
To monitor the ensuing dynamics we rely on the eGPE of Eq.~(\ref{eGPE}) using the additional imaginary term $- (i \hbar K_3 /2) \abs{\psi(\textbf{r},t)}^4 \psi(\vb{r},t)$, where $K_3$ refers to the three-body recombination rate~\cite{Bisset2015,Xi2016,Halder2022}. 
As a reference point of the recombination rate coefficient we consider $K_3=1.2\times10^{-40} m^6/s$ which was identified in the experiment of Ref.~\cite{Ferrier2016}. 
For a discussion about the interplay of three-body recombination and beyond mean-field contributions see the review in  Ref.~\cite{Chomaz2023}. 

The density evolution visualizing the emergent tunneling dynamics of the initially prepared SS in a tilted double-well for different energy offsets,  $D=10\hbar \omega_x/l_{{\rm osc}}$ and $D=7\hbar \omega_x/l_{{\rm osc}}$, is illustrated in Fig.~\ref{densities_1Dtunel_loss}(a) and (b) respectively. 
As it can be deduced, the presence of three-body losses does not prevent the observation of the tunneling processes taking place at early  timescales $t<200$ms that have been discussed in the main text, see also Figs.~\ref{densities_1Dtunel}(a2), (b2). 
Namely, for $D=10\hbar \omega_x/l_{{\rm osc}}$ oscillations of the SS between the wells occur. 
Also, in the case of $D=7\hbar \omega_x/l_{{\rm osc}}$ despite the overall collective tunneling behavior there are certain reflected portions of the SS simultaneously with each transmission event. In the course of the evolution they accumulate, and their interference leads to density smudging and locking of the tunneling process. 

However, as expected, the underlying atom losses depicted in Fig.~\ref{densities_1Dtunel_loss}(c) become more pronounced for longer times ($t \gg \omega_x^{-1}$) resulting ultimately in the destruction of the SS  because three-body losses compete with the LHY contribution. 
This can be readily seen in $n_{\rm 1D}(x,t)$ where the individual density humps gradually smoothen and disappear. 
The dBEC is not able to host a SS at these timescales due to the significant reduction of the atom number, $N(t)$. 
This decrease of the atom number is, of course, enhanced for larger $K_3$ coefficients as can also be inferred from Fig.~\ref{densities_1Dtunel_loss}(c).  However, it also depends on the initial energy offset, see Fig.~\ref{densities_1Dtunel_loss}(d), but apparently, there is a non-monotonic trend with respect to $D$. 
This behavior is caused by the peculiar dependence of the peak density on $D$, which is here the crucial component for the loss rate for fixed $K_3$. 
For instance, at $D=2\hbar \omega_x/l_{ { \rm osc}}$ where the individual crystals are self-trapped in each well, the density is more localized when compared to larger $D$ values as shown in the inset of Fig.~\ref{densities_1Dtunel_loss}(d). Therefore, the reduction of $N(t)$ is more rapid in this case. 
The same argument holds for the other $D$ values. 
A similar phenomenology occurs also for increasing $\epsilon_{{\rm dd}}$ (not shown), where droplet arrays form. In this case the atom losses are more pronounced due to the relatively higher localized densities related to negative chemical potentials, see also  Fig.~\ref{fig1_chem}(a).

\section{Linear ramping of the energy offset}\label{td_quench}

Another interesting question is whether the above-described tunneling regimes persist in the case of reducing the energy offset in a time-dependent manner or are unique to the quench protocol used. 
In the following, we argue that indeed the aforementioned tunneling regions can be realized utilizing time-dependent ramps with the latter being sufficiently fast, while closer to the diabatic regime the tunneling is naturally delayed. 
This investigation allows also to testify the validity of the eGPE predictions after quenching the tilt strength, a process that might induce excitations beyond the validity of this approach. 

\begin{figure}
\centering
\includegraphics[width = 0.49\textwidth]{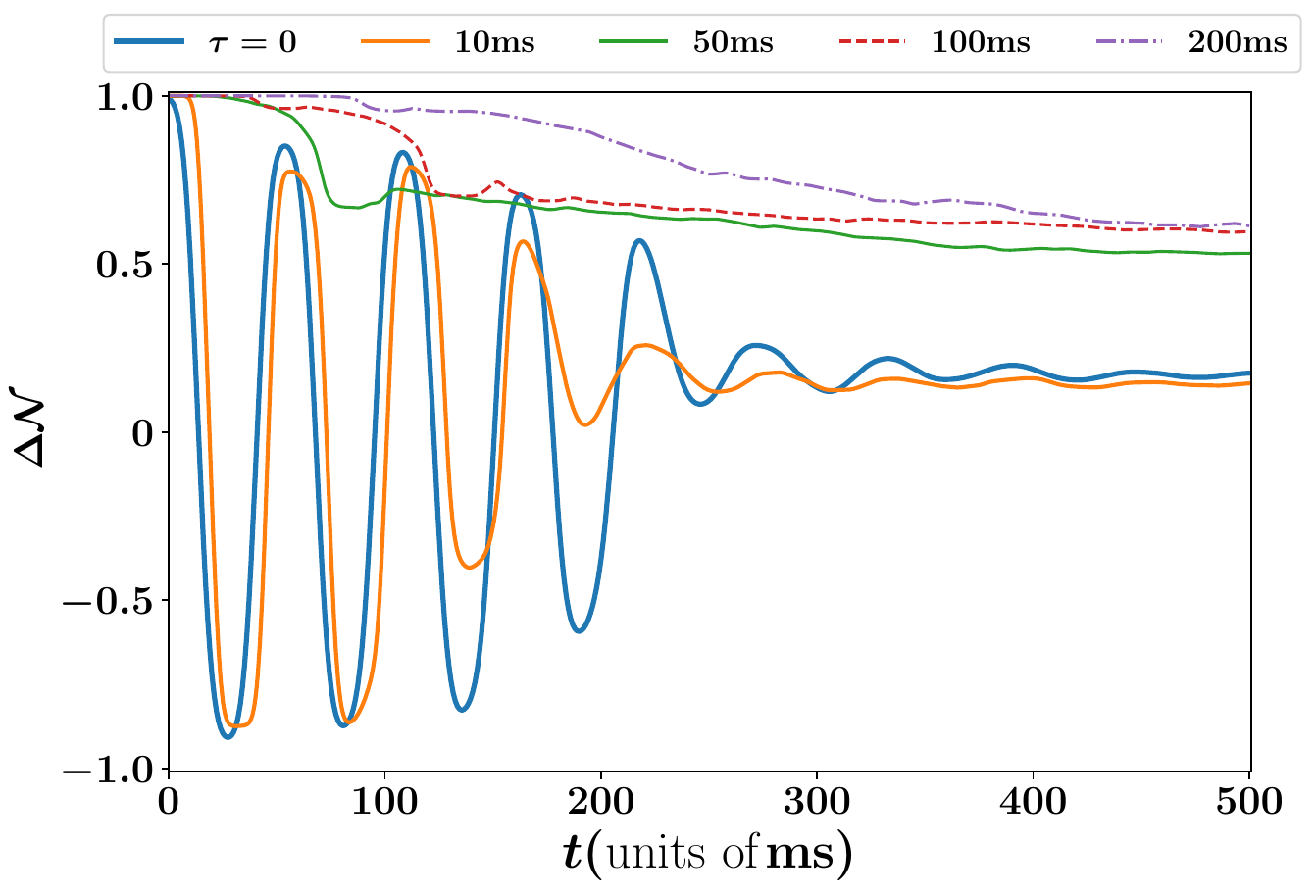}
\caption{Temporal evolution of the inter-well population imbalance, $\Delta \mathcal{N}(t)$, following a linear decrease of the energy offset from $D=7\hbar \omega_x/l_{{\rm osc}}$ to $D=0$ for different ramp-rates $\tau$ (see legend). 
A larger ramp-rate of the quench protocol leads progressively to tunneling suppression. 
The quasi-1D dBEC consists of $N=40000$ $^{164}$Dy atoms with $\epsilon_{{\rm dd}}=1.394$.} \label{populations_ramp_1D}
\end{figure}

For these reasons, instead of performing a quench of the energy offset, we also employ the linear ramp 
\begin{eqnarray}
&D(t;\tau)=D_{{\rm in}}+\frac{(D_{{\rm fi}}-D_{{\rm in}})t}{\tau}, {~~\rm for}~~t \leq \tau \\&
D(t;\tau)=D_{ {\rm fi}},~~~~~~~~~~~~~~~~~~~{\rm for}~~ t>\tau. 
\label{ramp_prot}
\end{eqnarray}
Here, $D_{{\rm in}}$ [$D_{{\rm fi}}$] denotes the initial [final] tilt strength and $\tau$ quantifies the underlying ramp rate. 
The latter is $\tau \to 0$ for a quench and $\tau \to \infty$ for a diabatic change of the energy offset. 
As in the main text we consider $D_{{\rm fi}}=0$ and for simplicity we restrict our presentation to the quasi-1D setting with $D_{{\rm in}}=7\hbar \omega_x/l_{{\rm osc}}$ and $\epsilon_{ {\rm dd}}=1.394$ referring to a SS.  
To monitor the inter-well tunneling features we rely on the population imbalance, $\Delta \mathcal{N}(t)$ in the course of the evolution showcased in Fig.~\ref{populations_ramp_1D}. 
Apparently, we observe that for sufficiently small ramp-rates, $\tau<10$ms$<\tau_{D_{{\rm in}}=7}$, the tunneling behavior is almost identical to the quench scenario, while a larger $\tau$ leads to a significantly slower evolution and eventually to suppression of tunneling, {\textit {i.e.}}, the magnetic atoms tend to remain in the initial well.

\section{Numerical Scheme}\label{numerics}

To numerically solve the 3D eGPE [Eq.~\eqref{eGPE}] we make use of the split-time Crank-Nicholson discretization scheme~\cite{CrankNicolson1947,Antoine2013} and deploy a suitable rescaling. 
In particular, we transform the 3D wave function as $\Psi(\vb{r'}, t') = \sqrt{l^3_{\rm{osc}}/N}\psi(\vb r, t)$ and express the time and length in units of the trap frequency $\omega_x$ and the harmonic oscillator length $l_{\rm osc} = \sqrt{\hbar/m \omega_x}$~, respectively. 
The ground-state of the dipolar gas is found via the imaginary time propagation method. 
Due to the many energetically close-lying configurations we employ various initial guesses in order to identify the many-body state with the lowest energy.  
Specifically, in the 1D case the following two different initial guesses $\psi_{1D}(x, y, z) = \mathcal{A}e^{-(x^2 + k^2 y^2+\lambda^2 z^2)/2}\cos^{2}(l_1x)$ and $\psi_{1D}(x, y, z) = \mathcal{A}e^{-(x^2 + k^2 y^2+\lambda^2 z^2)/2}\sin^2(l_1x)$ are used. 
The parameters $k^2=\omega_y/\omega_x$ and $\lambda^2=\omega_z/\omega_x$ stem from the above-mentioned rescaling.
On the other hand, for the 2D scenario we invoke the initial guesses $\psi_{2D}(x, y, z) = \mathcal{A}e^{-(x^2 + k^2 y^2+\lambda^{2} z^2)/2}(\cos^{2}(l_1x) + \cos^{2}(l_2 y + \phi))$ and $\psi_{2D}(x, y, z) = \mathcal{A}e^{-(x^2 + k^2 y^2+\lambda^2 z^2)/2}(\sin^{2}(l_1 x) + \cos^{2}(l_2  y + \phi))$. 
Here, $\mathcal{A}$ denotes the normalization constant while $l_1$, $l_2$, and $\phi$ are varied in order to realize a different number of crystals in the initial guess wave function. 
The choice of the initial ansatz is naturally more crucial in 2D due to the relatively higher degrees of freedom. 

To identify the correct ground state, we evaluate the corresponding energies of different initial states with precision up to ten decimal digits. 
The normalization of the wave function at every step of the imaginary time evolution is preserved by applying $\psi(\vb{r}', t) \rightarrow N^{1/2}/{\norm{\psi(\vb{r}', t)}}$. 
Convergence is justified by testifying that modifications of the wave function (at every grid point) between consecutive time-steps is lower than $10^{-6}$  and the corresponding energy alterations lie below $10^{-8}$. 
This serves as the initial state for the quench-induced tunneling dynamics which is monitored through real time propagation of the eGPE. 
The divergent behavior of the dipolar interaction potential at short distances is circumvented by transforming to momentum space~\cite{Goral2002} for the calculations. 
Subsequently, we employ the inverse Fourier transform to find the real space configurations relying on the convolution theorem. 
For the present simulations a 3D box with a spatial grid ($n_x \times n_y \times n_z$) is used which refers to ($1024\times128\times128$) in the quasi-1D case and ($512\times512\times128$) for the quasi-2D geometry. 
The spatial discretization is $\delta x = \delta y = \delta z =0.05~l_{{\rm osc}}$, and the time-step of the numerical integration $\delta t = 10^{-5}/\omega_x$. Hence, in order to accurately simulate the real time dynamics of the system we guarantee that $ ( \delta t )^2 < \delta x \delta y$ ~\cite{Antoine2013} is satisfied. 
Indeed, such spatial and time discretization steps ensure that the total particle number and total energy remain conserved, numerically of the order of $10^{-6}$, during the evolution time.

\bibliography{reference_ordered}	
\end{document}